\documentclass[runningheads]{llncs}
\usepackage{graphicx}
\usepackage{hyperref,xcolor}
\usepackage{comment}
\usepackage{algorithm,algpseudocode}
\usepackage{csquotes}
\begin{document}
%

\title{Pattern-based Acquisition of Scientific Entities from Scholarly Article Titles\thanks{Supported by TIB Leibniz Information Centre for Science and Technology, the EU H2020 ERC project ScienceGRaph (GA ID: 819536).}}

\titlerunning{Pattern-based Acquisition of Scientific Entities}
%
\author{Jennifer D'Souza\inst{1}\orcidID{0000-0002-6616-9509} \and
S\"oren Auer\inst{1,2}\orcidID{0000-0002-0698-2864}}
\authorrunning{D'Souza and Auer}
%
\institute{TIB Leibniz Information Centre for Science and Technology \and
L3S Research Center at Leibniz University of Hannover \\ Hannover, Germany \\
\email{\{jennifer.dsouza|auer\}@tib.eu}}
\maketitle              
\begin{abstract}

We describe a rule-based approach for the automatic acquisition of salient scientific entities from Computational Linguistics (CL) scholarly article titles. Two observations motivated the approach: (i) noting salient aspects of an article's contribution in its title; and (ii) pattern regularities capturing the salient terms that could be expressed in a set of rules. Only those lexico-syntactic patterns were selected that were easily recognizable, occurred frequently, and positionally indicated a scientific entity type. The rules were developed on a collection of 50,237 CL titles covering all articles in the ACL Anthology. In total, 19,799 \textit{research problems}, 18,111 \textit{solutions}, 20,033 \textit{resources}, 1,059 \textit{languages}, 6,878 \textit{tools}, and 21,687 \textit{methods} were extracted at an average precision of 75\%.


\end{abstract}

\section{Introduction}

Scientists increasingly face the information overload-and-drown problem even in narrow research fields given the ever-increasing flood of scientific publications~\cite{johnson2018stm,landhuis2016scientific}. Recently, solutions are being implemented in the domain of the digital libraries by transforming scholarly articles into ``digital-first'' applications as machine-interpretable scholarly knowledge graphs (SKGs), thus enabling completely new technological assistance to navigate the massive volumes of data through intelligent search and filter functions, and the integration of diverse analytics tools. There are several directions to this vision focused on representing, managing and linking metadata about articles, people, data and other relevant keyword-centered entities (e.g., Research Graph~\cite{aryani2018research}, Scholix~\cite{burton2017scholix}, Springer-Nature’s SciGraph or DataCite’s PID Graph~\cite{cousijn2021connected}, SemanticScholar~\cite{ammar2018construction}). This trend tells us that we are on the cusp of a great change in the digital technology applied to scholarly knowledge. Notably, next-generation scholarly digital library (DL) infrastructures have arrived: the Open Research Knowledge Graph (ORKG)~\cite{10.1145/3360901.3364435} digital research and innovation infrastructure by TIB and partner institutions, argues for obtaining a semantically rich, interlinked KG representations of the ``content'' of the scholarly articles, and, specifically, only \textit{research contributions}.\footnote{The ORKG platform can be accessed online: \url{https://orkg.org/} .} With intelligent analytics enabled over such contributions-focused SKGs, researchers can readily track research progress without the cognitive overhead that reading dozens of articles impose. A typical dilemma then with building such an SKG is deciding the type of information to be represented. In other words, what would be the information constituent candidates for an SKG that reflects the overview? While the scope of this question is vast, in this paper, we describe our approach designed with this question as the objective.

\textquote{\textit{Surprisingly useful information can be found with only a very simply understanding of the text.}}~\cite{hearst1992automatic} The quotation is the premise of the ``Hearst'' system of patterns which is a popular text mining method in the CL field. It implemented discovering lexical relations from a large-scale corpus simply by looking for the relations expressed in well-known ways. This simple but effective strategy was leveraged in supporting the building up of large lexicons for natural language processing~\cite{hearst1998automated}, e.g., the WordNet lexical project~\cite{miller1998wordnet}. Our approach is inspired after the ``Hearst'' methodology but on scholarly article titles content thereby implementing a pattern-based acquisition of scientific entities. Consider the two paper title examples depicted in Table 1. More fluent readers of English can phrase-chunk the titles based on lexico-syntactic patterns such as the colon punctuation in title 1 and prepositional phrase boundary markers (e.g., `to' in title 2). Following which, with some domain awareness, the terms can be semantically conceptualized or typed (e.g., as \textit{research problem}, \textit{resource}, \textit{method}, \textit{tool}, etc.). Based on such observations and circling back to the overarching objective of this work, we propose and implement a pattern-based acquisition approach to mine contribution-focused, i.e. salient, scientific entities from article titles. While there is no fixed notion of titles written with the purpose of reflecting an article's contribution, however, this is the generally known practice that it contains salient aspects related to the \textit{contribution} as a single-line summary. To the best of our knowledge, a corpus of only article titles remains as yet comprehensively unexplored as a resource for SKG building. Thus, our work sheds a unique and novel light on SKG construction representing \textit{research overviews}.

\begin{table}[!tb]
\centering
\begin{tabular}{|p{12cm}|}
\textbf{SemEval-2017 Task 5: Fine-Grained Sentiment Analysis on Financial Microblogs and News}\\ \textit{research\_problem}: {[}`SemEval-2017 Task 5'{]}\\ \textit{resource}: {[}`Financial Microblogs and News'{]}\\ \textit{method}: {[}`Fine-Grained Sentiment Analysis'{]}  \\ \\


\textbf{Adding Pronunciation Information to Wordnets}\\ \textit{solution}: {[}'Adding Pronunciation Information'{]}\\ \textit{tool}: {[}`Wordnets'{]} \\ \\

\end{tabular}
\caption{Two examples of scholarly article titles with their concept-typed scientific terms which constitutes the IE objective of the \textsc{CL-Titles-Parser}}
\end{table}

In this paper, we discuss \textsc{CL-Titles-Parser} -- a tool for extracting salient scientific entities based on a set of lexico-syntactic patterns from titles in Computational Linguistics (CL) articles. 
Six concept types of entities were identified applicable in CL titles, viz. \textit{research problem}, \textit{solution}, \textit{resource}, \textit{language}, \textit{tool}, and \textit{method}. \textsc{CL-Titles-Parser} when evaluated on almost all titles (50,237 of 60,621 total titles) in the ACL Anthology performs at a cumulative average of 75\% IE precision for the six concepts. Thus, its resulting high-precision SKG integrated in the ORKG can become a reliable and essential part of the scientist's workbench in visualizing the overview of a field or even as crowdsourcing signals for authors to describe their papers further. \textsc{CL-Titles-Parser} is released as a standalone program \url{https://github.com/jd-coderepos/cl-titles-parser}.


\section{Related Work}

\paragraph{\textbf{Key Summary of Research in Phrasal Granularity}} 
To bolster search technology, the phrasal granularity was used to structure the scholarly record. Thus scientific phrase-based entity annotated datasets in various domains including multidisciplinarily across STEM~\cite{handschuh2014acl,augenstein2017semeval,Luan2018MultiTaskIO,lrec2020} were released; machine learning systems were also developed for automatic scientific entity extraction~\cite{Ammar2017TheAS,luan2017scientific,Beltagy2019SciBERTPC,brack2020domain}. 
However, none of these resources are clearly indicative of capturing only the salient terms about research contributions which is the aim of our work. 

\paragraph{\textbf{Pattern-based Scientific Terminology Extraction}} 
Some systems~\cite{heffernan2018identifying} viewed key scholarly information candidates as problem-solution mentions. \cite{charles2011adverbials} used the discourse markers ``thus, therefore, then, hence'' as signals of problem-solution patterns. 
\cite{gupta2011analyzing} used semantic extraction patterns learned via bootstrapping to the dependency trees of sentences in Abstracts to mine the research focus, methods used, and domain problems. 
Houngbo and Mercer~\cite{houngbo2012method} extracted the methods and techniques from biomedical papers by leveraging regular expressions for phrase suffixes as ``algorithm,'' ``technique,'' ``analysis,'' ``approach,'' and ``method.'' AppTechMiner~\cite{singh2017apptechminer} used rules to extract application areas and problem solving techniques. The notion of application areas in their model is analogous to research problem in ours, and their techniques are our tool or method. Further, their system extracts research problems from the article titles via rules based on functional keywords, such as, ``for,'' ``via,'' ``using'' and ``with'' that act as delimiters for such phrases. \textsc{CL-Titles-Parser} also extracts problems from titles but it does so in conjunction with other information types such as tools or methods. AppTechMiner uses citation information to determine term saliency. In contrast, since we parse titles, our data source itself is indicative of the saliency of the scientific terms therein w.r.t. the article's contribution. Finally, \cite{katsurai2021adoption}, like us, use a system of patterns to extract methods from the titles and Abstracts of articles in Library Science research. We differ in that we extract six different types of scientific entities and we focus only on the article titles data source.

Next, in the article, we describe the \textsc{CL-Titles-Parser} for its pattern-based acquisition of scientific entities from Computational Linguistics article titles.

\section{Preliminary Definitions}

We define the six scientific concept types handled in this work. The main aim here is not to provide rigorous definitions, but rather just to outline essential features of the concepts to explain the hypotheses concerning their annotation.

\textit{\textbf{i. Research problem.}} The theme of the investigation. E.g., ``Natural language inference.'' In other words, the answer to the question ``which problem does the paper address?'' or ``On what topic is the investigation?'' \textit{\textbf{ii. Resource.}} Names of existing data and other references to utilities like the Web, Encyclopedia, etc., used to address the \textit{research problem} or used in the \textit{solution}. E.g., ``Using Encyclopedic Knowledge for Automatic Topic Identification.'' In this sentence, ``Encyclopedic Knowledge'' is a \textit{resource} used for \textit{research problem} ``Automatic Topic Identification.'' \textit{\textbf{iii. Tool.}} Entities arrived at by asking the question ``Using what?'' or ``By which means?'' A \textit{tool} can be seen as a type of a \textit{resource} and specifically software. \textit{\textbf{iv. Solution.}} A novel contribution of a work that solves the \textit{research problem}. E.g., from the title ``PHINC: A Parallel Hinglish Social Media Code-Mixed Corpus for Machine Translation,'' the terms `PHINC' and `A Parallel Hinglish Social Media Code-Mixed Corpus' are solutions for the \textit{problem} `Machine Translation.' \textit{\textbf{v. Language.}} The natural language focus of a work. E.g., Breton, Erzya, Lakota, etc. \textit{Language} is a pertinent concept w.r.t. an overview SKG about NLP solutions. \textit{\textbf{vi. Method.}} They refer to existing protocols used to support the \textit{solution}; found by asking ``How?''

\section{Tool Description}

\subsection{Formalism}

Every CL title $T$ can be expressed as one or more of the following six elements $te_i = \langle rp_i, \allowbreak res_i, \allowbreak tool_i, \allowbreak lang_i, \allowbreak sol_i, \allowbreak meth_i \rangle$, representing the \textit{research problem}, \textit{resource}, \textit{tool}, \textit{language}, \textit{solution}, and \textit{method} concepts, respectively. A title can contain terms for zero or more of any of the concepts. 
The goal of \textsc{CL-Titles-Parser} is, for every title $t_i$, to annotate its title expression $te_i$, involving scientific term extraction and term concept typing.

\subsection{Rule-based Processing Workflow}

\textsc{CL-Titles-Parser} operates in a two-step workflow. First, it aggregates titles as eight main template types with a default ninth category for titles that could not be clustered by any of the eight templates. Second, the titles are phrase-chunked and concept-typed based on specific lexico-syntactic patterns that are group-specific. The concept type is selected based on the template type category and some contextual information surrounding the terms such as prepositional and verb phrase boundary markers. 

\paragraph{Step 1: Titles clustering based on commonly shared title lexico-syntactic patterns.} While our rule-based system implements eight patterns in total, we describe four template patterns as examples.


\textbf{Template ``hasSpecialCaseWord() :''} applies to titles written in two parts -- a one-word solution name, a colon separator, and an elaboration of the solution name. E.g., ``SNOPAR: A Grammar Testing System'' consisting of the one word `SNOPAR' solution name and its elaboration `A Grammar Testing System.' Further, there are other instances of titles belonging to this template type that are complex sentences, i.e. titles with additional prepositional or verb phrases, where mentions of the \textit{research problem}, \textit{tool}, \textit{method}, \textit{language} domain etc. are also included in the latter part of the title. E.g., ``GRAFON: A Grapheme-to-Phoneme Conversion System for Dutch'' is a complex title with a prepositional phrase triggered by ``for'' specifying the \textit{language} domain ``Dutch.'' 

\textbf{Template ``Using ...''} applies to titles that begin with the word ``Using'' followed by a \textit{resource} or \textit{tool} or \textit{method} and used for the purpose of a \textit{research problem} or a \textit{solution}. E.g., the title ``Using WordNet for Building WordNets'' with \textit{resource} ``WordNet'' for \textit{solution} ``Building WordNets''; or ``Using Multiple Knowledge Sources for Word Sense Discrimination'' with \textit{resource} ``Multiple Knowledge Sources'' for \textit{research problem} ``Word Sense Discrimination.''

\textbf{Template ``... case study ...''} Titles in this category entail splitting the phrase on either side of ``case study.'' The first part is processed by the precedence-ordered rules to determine the concept type. The second part, however, is directly cast as \textit{research problem} or \textit{language} since they were observed as one of the two. The checks for \textit{research problem} or \textit{language} are made by means of regular expressions implemented in helper functions. E.g., the title ``Finite-state Description of Semitic Morphology: A Case Study of Ancient Accadian'' would be split as ``Finite-state Description of Semitic Morphology'' and ``Ancient Accadian'' language domain, where ``Ancient Accadian'' is typed as \textit{language} based on regex patterns. See Table 2 for examples of some regular expressions employed. 

\begin{table}[!htb]
\centering
\begin{tabular}{ll} \hline
\textit{languages} & $reLanguage = (...|Tigrigna|\allowbreak Sundanese|\allowbreak Balinese|...)$ \\
\textit{tool}      & $reTool = (...|memory|\allowbreak controller|\allowbreak workbench(es)?|...)$ \\
\textit{resource}      & $reResource = (...|corp(ora|us)|\allowbreak vocabular(ies|y)|\allowbreak cloud|...)$ \\
\textit{method} & $reMethod = (...|protocol|\allowbreak methodolog(ies|y)|\allowbreak recipe|...)$ \\ \hline
\end{tabular}
\caption{Regular expressions in suffix patterns for scientific term concept typing}
\end{table}

\textbf{Template ``... : ...''} A specialized version of this template is ``hasSpecialCaseWord() :''. Here, titles with two or more words in the phrase preceding the colon are considered. They are split in two parts around the colon. The parts are then further processed to extract the scientific terms. E.g., ``Working on the Italian Machine Dictionary: A Semantic Approach'' split as ``Working on the Italian Machine Dictionary'' and ``A Semantic Approach'' where the second part is a non-scientific information content phrase. By non-scientific information content phrase we mean phrases that cannot be categorized as one of the six concepts. 

\paragraph{Step 2: Precedence-ordered scientific term extraction and typing rules.} The step is conceptually akin to sieve-based systems that were successfully demonstrated on the coreference resolution~\cite{raghunathan2010multi} and biomedical name normalization~\cite{d2015sieve} tasks. The idea in sieve-based systems is simply that an ordering is imposed on a set of selection functions from the most constrained to the least. Similarly, in this second step of processing titles in our rule-based system, we apply the notion of selection precedence as concept precedence. However, there are various concept precedences in our system that depend on context information seen as the count of the connectors and the type of connectors in any given article title. 

In this step, within each template category the titles are generally processed as follows. \textbf{Step 1. Counting connectors} -- Our connectors are a collection of 11 prepositions and 1 verb defined as: $connectors\_rx = (to|\allowbreak of|\allowbreak on|\allowbreak for|\allowbreak from|\allowbreak with|\allowbreak by|\allowbreak via|\allowbreak through|\allowbreak using|\allowbreak in|\allowbreak as)$. For a given title, its connectors are counted and the titles are phrase chunked as scientific phrase candidates split on the connectors themselves. \textbf{Step 2. Concept typing} -- This involves selecting the workflow for typing the scientific phrases with concept types among our six, viz. \textit{language}, \textit{tool}, \textit{method}, \textit{resource}, and \textit{research problem}, based on context information. Workflow branches were implemented as a specialized system of rules based on the number of connectors. The next natural question is: after the workflow branch is determined, what are the implementation specifics for typing the scientific terms per our six concepts? We explain this with the following specific case. A phrase with 0 connectors is typed after the following concept precedence order: $\textit{language} \prec \textit{tool} \prec \textit{method} \prec \textit{resource} \prec \textit{research problem}$ where each of the concepts are implemented as regex checks. Some example regexes were shown earlier in Table 2. And it only applies to five of the six concepts, i.e. \textit{solution} is omitted. Alg. 1 pseudocode in Appendix~\ref{no-connectors-alg} illustrates this.\footnote{Pseudocodes are offered as supplementary content in this paper as they are mere translations of our available code \url{https://github.com/jd-coderepos/cl-titles-parser}.} On the other hand, if a title has one connector, it enters first into the OneConnectorHeu() branch (see Alg.~\ref{alg:one-connector-heuristics} in Appendix~\ref{one-connector}). There, the first step is determining which connector is in the phrase. Then based on the connector, separate sets of concept type precedence rules apply. The concept typing precedence rules are tailored based on the connector context. For instance, if the connector is `from,' the title subphrases are typed based on the following pattern: \textit{solution} from \textit{resource}.

This concludes a brief description of the working of \textsc{CL-Titles-Parser}.

\section{Evaluation}

In this section, some results from \textsc{CL-Titles-Parser} are discussed for scientific term extraction and concept typing in Computational Linguistics article titles.

\paragraph{\textbf{Evaluation Corpus}} We downloaded all the article titles in the ACL anthology as the `Full Anthology as BibTeX' file dated 1-02-2021. See \url{https://aclanthology.org/anthology.bib.gz}. From a total of 60,621 titles, the evaluation corpus comprised 50,237 titles after eliminating duplicates and invalid titles.

When applied to the evaluation corpus, the following total scientific concepts were extracted by the tool: 19,799 \textit{research problem}, 18,111 \textit{solution}, 20,033 \textit{resource}, 1,059 \textit{language}, 6,878 \textit{tool}, and 21,687 \textit{method}. These scientific concept lists were then evaluated for extraction precision.

\subsection{Quantitative Analysis: Scientific Concept Extraction Precision}

First, each of the six scientific concept lists were manually curated by a human annotator to create the gold-standard data. The extracted lists and the gold-standard lists were then evaluated w.r.t. the $precision$ metric. Table~\ref{tab:maneval} shows the results. We see that \textsc{CL-Titles-Parser} demonstrates a high information extraction precision for all concept types except \textit{research problem}. This can be attributed in part to the long-tailed list phenomenon prevalent in the scientific community as the scholarly knowledge investigations steadily progress. With this in mind, the gold-standard list curation was biased toward already familiar research problems or their derivations. Thus we estimated that at least 20\% of the terms were pruned in the gold data because they were relatively new as opposed to being incorrect. Note, recall evaluations were not possible as there is no closed-class gold standard as scientific terms are continuously introduced.

\begin{table}[!b]
\centering
\begin{tabular}{ll|ll}
 Concept type   & $Precision$ & Concept type & $Precision$ \\ \hline
\textit{research problem}  &  58.09\%   & \textit{method} & 77.29\% \\
\textit{solution} &  80.77\% & \textit{language} & 95.12\%   \\
\textit{tool} & 83.40\%  & \textit{resource} & 86.96\%
\end{tabular}
\caption{Precision of \textsc{CL-Titles-Parser} for scientific term extraction and concept typing from 50,237 titles in the ACL anthology}
\label{tab:maneval}
\end{table}

\subsection{Qualitative Analysis: Top N Terms}

As qualitative analysis, we examine whether the terms extracted by our tool reflect popular research trends. \autoref{top_5} shows the top five terms in each of the six concept types. The full scientific concept lists sorted by occurrences are available in our code repository \url{https://github.com/jd-coderepos/cl-titles-parser/tree/master/data-analysis}. Considering the \textit{research problem} concept, we see that variants of ``machine translation'' surfaced to the top accurately reflective of the large NLP subcommunity attempting this problem. As a \textit{tool}, ``Word Embeddings'' are the most predominantly used. ``Machine Translation'' itself was the most employed \textit{method}. Note that the concept types are not mutually exclusive. A term that is a \textit{research problem} in one context can be a \textit{method} in a different context. As an expected result, ``English'' is the predominant \textit{language} researched. ``Twitter'' showed as the most frequently used \textit{resource}. Finally, predominant \textit{solutions} reflected the nature of the article itself as ``overview'' or ``a study'' etc. Then \autoref{top_5_century} shows the \textit{research problem}, \textit{resource}, and \textit{tool} concepts research trends in the 20th vs. 21st century. Contemporarily, we see new predominant neural \textit{research problem} mentions, an increasing use of social media as a \textit{resource}; and various neural networks as \textit{tools}.

\begin{table}[!tb]
\centering
\begin{tabular}{|p{1.3cm}|p{10.7cm}|} \hline
\textit{research-problem} & statistical machine translation (267), machine translation (266), neural machine translation (193), sentiment analysis (99), information extraction (85) \\ \hline
\textit{tool} & word embeddings (77), neural networks (63), conditional random fields (51), convolutional neural networks (41), spoken dialogue systems (32) \\ \hline
\textit{method} & machine translation (105), domain adaptation (68), sentiment analysis (68), named entity recognition (67), statistical machine translation (66) \\ \hline
\textit{language} & English (150), Chinese (87), Japanese (87), German (81), Arabic (74) \\ \hline
\textit{resource} & Twitter (204), text (173), social media (132), the web (115), Wikipedia (98) \\ \hline
\textit{solution} & overview (39), a study (23), an empirical study (25), a comparison (21), a toolkit (17) \\ \hline
\end{tabular}
\caption{Top 5 scientific phrases for the six concepts extracted by \textsc{CL-Titles-Parser}}
\label{top_5}
\end{table}

\begin{table}[!tb]
\begin{tabular}{l|p{4cm}|p{2.3cm}|p{4.9cm}|} \cline{2-4}
     & \textit{research problem} & \textit{resource} & \textit{tool} \\ \cline{2-4}
20th & machine translation (56) & text (38)  & machine translation system (8)  \\
21st & statistical machine translation (258) & text (251) & word embeddings (87) \\ \hline
20th & information extraction (19) & discourse (17) & natural language interfaces (7) \\
21st & machine translation (210) & Twitter (204) & neural networks (57) \\ \hline
20th & speech recognition (16) & TAGs (9)  & neural networks (6) \\
21st & neural machine translation (193) & social media (132) & conditional random fields (51) \\ \hline
20th & natural language generation (15) & bilingual corpora (9) & WordNet (3) \\
21st & sentiment analysis (99) & the web (115) & convolutional neural networks (41) \\ \hline
20th & continuous speech recognition (12) & dialogues (9) & semantic networks (3) \\
21st & question answering (81) & Wikipedia (98) & spoken dialogue systems (31) \\ \cline{2-4}
\end{tabular}
\caption{Top 5 \textit{research problem}, \textit{resource}, and \textit{tool} phrases from paper titles reflecting research trends in the 20th (7,468 titles) vs. the 21st (63,863 titles) centuries.}
\label{top_5_century}
\end{table}

\section{Conclusion and Future Directions}

We have described a low-cost approach for automatic acquisition of contribution-focused scientific terms from unstructured scholarly text, specifically from Computational Linguistics article titles. Work to extend the tool to parse Computer Science titles at large is currently underway. The absence of inter-annotator agreement scores to determine the reliability with which the concepts can be selected will also be addressed in future work. Evaluations on the ACL anthology titles shows that our rules operate at a high precision for extracting \textit{research problem}, \textit{solution}, \textit{resource}, \textit{language}, \textit{tool}, and \textit{method}. We proposed an incremental step toward the larger goal of generating contributions-focused SKGs.

\bibliographystyle{splncs04}
\bibliography{cite}

\appendix

\section{Additional Examples}
In this section, we offer additional supplementary examples for Step 1 in \textsc{CL-Titles-Parser}. Considering the \textbf{``hasSpecialCaseWord() :''} template, the title ``CIRCSIM-Tutor: An Intelligent Tutoring System Using Natural Language Dialogue'' is an example of a complex title owing to the verb ``Using'' which makes it a multi-phrase syntactic structure. This title contains the brief solution name ``CIRCSIM-Tutor,'' a descriptive elaboration of the solution name ``An Intelligent Tutoring System,'' and a mention of a \textit{resource} ``Natural Language Dialogue.'' Another complex title is ``MDWOZ: A Wizard of Oz Environment for Dialog Systems Development.'' It contains the brief solution name ``MDWOZ,'' a descriptive elaboration of the solution name ``A Wizard of Oz Environment,'' and a \textit{research problem} ``Dialog Systems Development.''

Next, for the \textbf{``... : ...''} template. Examples of titles in this type were more varied, nonetheless, surface patterns could still be seen. ``The Lincoln Continuous Speech Recognition System: Recent Developments and Results'' is an example of a title where its second part, i.e. ``Recent Developments and Results'' is a non-scientific information content phrase per our six scientific concepts. Nevertheless, there were phrases where both parts could be processed. Consider the titles, ``The Problem of Naming Shapes: Vision-Language Interface,'' and ``Specialized Information Extraction: Automatic Chemical Reaction Coding From English Descriptions'' as representative examples.

\section{Coverage of the Rules}
In Section 4.2, in the article, we discussed the first titles clustering Step 1 that grouped article titles based on commonly used lexico-syntactic patterns. \autoref{tab:maneval} offers a view of which patterns were most predominant in the titles.

\begin{table}[!htb]
\centering
\begin{tabular}{lll|lll}
&  Rule   & $Total\_parsed$ & & Rule & $Total\_parsed$ \\ \hline
1 & \textbf{\textit{hasSpecialCaseWord() :}} & 3403 & 5 & contains \textbf{\textit{:}} & 8739 \\
2 & \textbf{\textit{Using ...}} & 915 & 6 & contains \textbf{\textit{applied to}} & 35 \\
3 & \textbf{\textit{... : [a ]?case study}} & 347 & 7 & \textbf{\textit{nonContentPhrase()}} & 1652 \\
4 & contains \textbf{\textit{[a ]?case study}} & 20 & 8 & \textbf{\textit{Description of ...}} & 18 \\
\end{tabular}
\caption{\textsc{CL-Titles-Parser} per lexicosyntactic pattern grouped titles in Step 1 from 50,237 total evaluated titles. \footnotesize{The default rule 9 parsed 34,926 titles.}}
\label{tab:maneval}
\end{table}

\section{Quantitative Analysis: Research Problem Recall}

Leveraging public data sources, we could at least create oracles for existing \textit{research problems}. Thus, this concept could be supplementarily evaluated for its recall. We performed two separate evaluations: 1) using lists from Wikipedia (\url{https://en.wikipedia.org/wiki/Natural_language_processing}, \url{https://en.wikipedia.org/wiki/Outline_of_natural_language_processing#Subfields_of_natural_language_processing}); and 2) using predefined lists from the web (\url{https://github.com/Kyubyong/nlp_tasks}) excluding Wikipedia. The results are reported in \autoref{recall}. Note that these results may be affected by the fact that the research problems in the predefined lists may not have been present in any of the article titles in the ACL anthology. Thus these results are only as an estimate of our tool's ability to identify \textit{research problems} which depends in a large part on the underlying corpus. 

\begin{table}[!htb]
\centering
\begin{tabular}{l|l|l}
       & \textit{Wikipedia} & \textit{Web} \\ \hline
\textit{research problem} recall & 61.33\%  & 75.91\%
\end{tabular}
\caption{The \textsc{CL-Titles-Parser} recall for extracting \textit{research problems} from the ACL Anthology compared with predefined lists from Wikipedia and the Web}
\label{recall}
\end{table}

\section{Precedence-Ordered Scientific Term Extraction for Phrases without Connectors}
\label{no-connectors-alg}

\begin{algorithm}[htb]
\caption{Five-way precedence-ordered concept typing}
\label{alg:concept-typing}
\begin{algorithmic}[1]
\State $ph$: a phrase extracted from an article title
    \Procedure{FiveWayConceptTyping}{$ph$}
    \State Initialize $lang, tool, meth, res, meth, rp = []$
    \If{is\_language(ph)}
        \State lang.append(ph)
    \ElsIf{is\_tool(ph)}
        \State tool.append(ph)
    \ElsIf{is\_method(ph)}
        \State meth.append(ph)
    \ElsIf{is\_resource(ph)}
        \State res.append(ph)
    \ElsIf{is\_research\_problem(ph)}
        \State rp.append(ph)
    \EndIf
    \State \textbf{return} $\langle lang, tool, meth, res, meth, rp \rangle$
    \EndProcedure
\end{algorithmic}
\end{algorithm}

\section{Miscellaneous Helper Function}

Algorithm 2 is an example helper function, that does not determine a scientific phrase concept but are leveraged as utilities in other functions in \textsc{CL-Titles-Parser}. 

\begin{algorithm}
\small
\caption{Phrase ending check with helper regex}
\label{alg:tool-concept-typing}
\begin{algorithmic}[1]
\State $ph$: a phrase extracted from an article title
\State $rx$: the helper regular expression
    \Procedure{Ending}{$ph,rx$}
    \State \textbf{return} not re.match('\^{}.*?('+rx+')\$', phrase) is None
    \EndProcedure
\end{algorithmic}
\end{algorithm}

\section{Processing of Titles with One Connector}
\label{one-connector}

In Step 2 of the rule-based processing workflow described in Section 4.2 in the main article, we discussed the ``OneConnectorHeu()'' workflow branch which is illustrated for its pseudocode in this section. We refer the reader to our released code for the implementation and further experimentation details \url{https://github.com/jd-coderepos/cl-titles-parser}.

\begin{algorithm}[!t]
\caption{One connector heuristics}
\label{alg:one-connector-heuristics}
\begin{algorithmic}[1]
\State $ph$: a phrase extracted from an article title
    \Procedure{OneConnectorHeuristics}{$ph$}
    \State Initialize $ph_lower = ph.lower()$
    \State $sol, rp, res, lang, tool, meth = []$
    \If{hasOnlyConnector(ph\_lower, for)}
        \State sol, rp, res, lang = forConnHeu(ph)
    \ElsIf{hasOnlyConnector(ph\_lower, of)}
        \State sol, rp, res, lang, tool = ofConnHeu(ph)
    \ElsIf{hasOnlyConnector(ph\_lower, `using|with|by')}
        \State sol, res, lang, tool, meth = usingWithByConnHeu(ph)
    \ElsIf{hasOnlyConnector(ph\_lower, on)}
        \State sol, rp, res, lang = onConnHeu(ph)
    \ElsIf{hasOnlyConnector(ph\_lower, from)}
        \State sol, res = fromConnHeu(ph)
    \ElsIf{hasOnlyConnector(ph\_lower, in)}
        \State res, rp, sol, lang, tool = inConnHeu(ph)
    \ElsIf{hasOnlyConnector(ph\_lower, `through|via')}
        \State sol, rp, meth, res = thruViaConnHeu(ph)
    \ElsIf{hasOnlyConnector(ph\_lower, to)}
        \State res, rp, sol, lang, tool, meth = toConnHeu(ph)
    \ElsIf{hasOnlyConnector(ph\_lower, as)}
        \State res, rp, sol, meth = asConnHeu(ph)        
    \EndIf
    \State \textbf{return} $\langle sol, rp, res, lang, tool, meth \rangle$
    \EndProcedure
\end{algorithmic}
\end{algorithm}

\end{document}